\documentclass[prb,twocolumn, superscriptaddress, floatfix]{revtex4}
\usepackage{graphicx}

\begin{document}

\title{The role of the band gaps in reconstruction of polar surfaces and interfaces}

\author{X. Gu}
\affiliation{Advanced Materials and Process Engineering Laboratory,
University of British Columbia, Vancouver, BC V6T 1Z4, Canada}

\author{I.S. Elfimov}
\affiliation{Advanced Materials and Process Engineering Laboratory, 
University of British Columbia, Vancouver, BC V6T 1Z4, Canada}

\author{G.A. Sawatzky}
\affiliation{Advanced Materials and Process Engineering Laboratory, 
University of British Columbia, Vancouver, BC V6T 1Z4, Canada}
\affiliation{Department of Physics and Astronomy, University of 
British Columbia, Vancouver, BC V6T-1Z1, Canada}

\begin{abstract}
Density functional theory applied to a simple ionic material, MgO, is
used as a model system to clarify several aspects of electronic driven
mechanism to compensate for the diverging electrostatic potential in
the polar structures. We demonstrate that in the stoichiometric free
standing films, the electronic reconstruction is limited by the band
gap. This produces a residual electric field in the bulk of the sample
which is extremely sensitive to tiny deviations in electron transfer
between two surfaces of the slab. In heterostructures, the band gap is
replaced by a new effective energy scale set by the band alignment of
its component. This changes the reconstruction pathways so that the
electronic mechanism can benefit from the smallest energy scale
possible.
\end{abstract}

\maketitle

Beyond a doubt, the electrostatic potential plays a fundamental role 
in the surface and bulk properties of virtually any ionic material. 
It has been long recognized as the driving force behind the so-called 
polar catastrophe that controls the physics and chemistry of polar 
surfaces and interfaces
\cite{BERTAUT:1958p3447,Lacman:1965p195,Tasker:1979p15,Wolf:1992p16}. 
A prominent example is a discovery of a 2D electron gas at the 
interface between LaAlO$_3$ and SrTiO$_3$\cite{Ohtomo:2004p423}. 
In the bulk, each one of these perovskites is an insulator with large 
conductivity gap of 5.6 and 3.25eV, respectively. However, when the 
two are combined in a heterostructure such that the (001) plane 
becomes an interface, a new metallic state is 
created\cite{Ohtomo:2004p423}. 
Several proposals have been put forward to explain the effect. 
It is argued that the large sheet charge density could be a result 
of Oxygen vacancies formed in SrTiO$_3$ during the growth process 
followed by the spatial separation between the charge compensating 
electrons and the vacancies\cite{Siemons:2007p196802}. 
Yoshimatsu {\it et. al.} suggested the conventional band bending 
mechanism based on the observation of the core level shifts seen 
in photoelectron spectroscopy for LaO-TiO$_2$ interface in a soft 
x-ray regime\cite{Yoshimatsu:2008p26802}.  
The mechanism of particular interest is that of electronic 
reconstruction\cite{Hesper:2000p16046}  resulting in self doping of 
the interface layers with carrier density and mobility that are 
much higher than currently achieved in conventional semiconductors.
The feasibility of electronic reconstruction scenario is currently 
under extensive investigation from experimental as well as 
theoretical standpoint
\cite{Okamoto:2004p630,Nakagawa:2006p204,Lee:2007p075339,
Wadati:2008arXiv:0810.1926v1,Sing:2009p176805,Salluzzo:2009p166804,
Savoia:2009p075110,Takizawa:2009p236401,Pentcheva:2009p107602}.
Here we have examined  the role played by the band structure in 
electronic reconstruction of polar surfaces and interfaces. 
To avoid unnecessary complexity, we focus on simple materials 
keeping all other degrees of freedom frozen. We chose magnesium 
oxide because of simple crystal structure and high ionicity even 
though we do not expect MgO to electronically reconstruct because 
of the large cost in energy due to the large band gap.

\begin{figure}
\includegraphics[clip=true,width=0.45\textwidth]{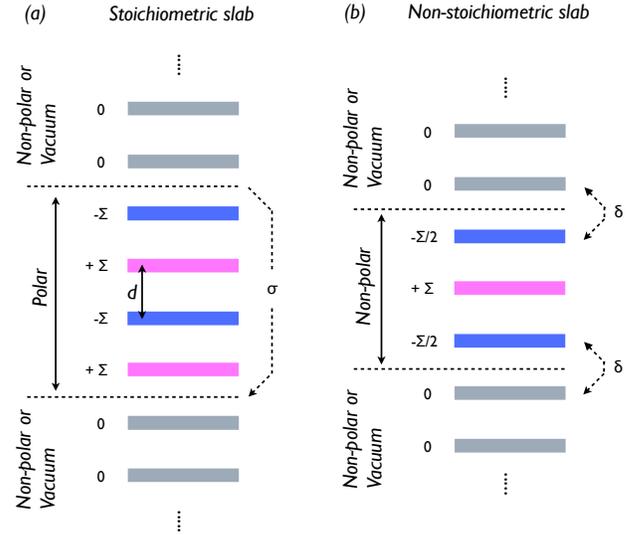}
\caption {An artist's concept of charge neutral stoichiometric 
(a) and non-stoichiometric (b) slabs. $\Sigma$ is a planar 
charge density ($Q/A$). The arrows depict electron transfer 
due to electronic reconstruction ($\sigma$) and charge density 
redistribution at the interface ($\delta$) due to the band 
structure effects.}
\label{geometry}
\end{figure}

\begin{figure}
\includegraphics[clip=true,width=0.45\textwidth]{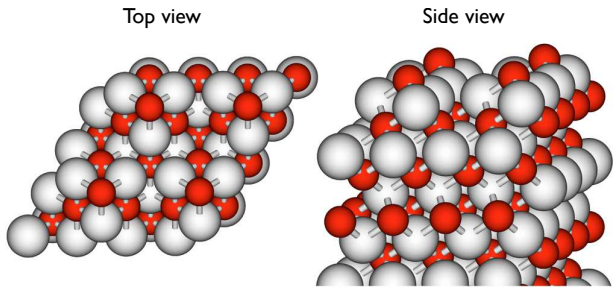}
\caption {Schematic representation of the octopolar 
reconstruction of polar (111) surface of an ionic material 
with rock-salt structure.}
\label{MgO111-octopolar}
\end{figure}

According to Tasker's classification scheme, the (001) surface 
of LaAlO$_3$ and (111) surface of MgO belong to type 3 polar 
termination of an ionic crystal and thus should not exist 
without a major reconstruction\cite{Tasker:1979p15}. 
The instability occurs due to accumulation of the electrostatic 
potential in the film in the direction parallel to the surface 
normal. Each unit cell contributes the same potential drop 
resulting in the electrostatic potential diverging with the 
film thickness. By virtue of its construction, only stoichiometric 
films with unalike polar terminations (Fig\ref{geometry}(a)) 
exhibit a divergent potential inside the material which drives 
the surface/interface reconstruction. In the non-stoichiometric 
structures with both film surfaces alike, the surface electronic 
structure is governed by the requirement of overall charge 
neutrality. For example, a defect-free (111) MgO film terminated 
on both sides with Mg is expected to be metallic in band theory 
because there is not enough Oxygen in the system to fully 
oxidize the Magnesium. Given the geometry of the structure and 
the Coulomb repulsion between alike charges, it is clear that 
the Mg on this surface should be 1+ just like the Mg on the 
surface of a stoichiometric (111) MgO slab in the case of 
electronic reconstruction. This approach kind of bypasses 
rather than treats the polar catastrophe issue but is a convenient 
approximation to get insight into the properties of excess 
charge density at each interface separately that might not 
be feasible otherwise\cite{Okamoto:2004p630,Pentcheva:2006p47}. 
Note that the term non-stoichiometric is used by us in a rather 
specific sense and refers to the systems in which the number of 
positive and negative charged atomic layers differs by one. 
It does not include formation of point defects such as vacancies 
nor adsorption of the foreign species on the surface.

Unlike type 3, the non-polar crystal truncations are a single 
boundary problem. Here, the properties are determined by the 
discontinuity in the crystal potential at a single interface 
with another material or vacuum. For example, the Shockley 
surface state on (111) surfaces of simple metals such as Cu, 
Ag or Au is a result of such discontinuity\cite{Shockley:1939p718}. 
In a case of charge neutral (type 1) surfaces of the ionic 
materials, it is the disruption of the Madelung potential
and the change in the local symmetry and coordination numbers 
that governs a change in the electronic structure of a material 
at the surface. Some layered structures such as realized in 
TiS$_2$ where Ti$^{4+}$ layer is sandwiched between two S$^{2-}$ 
layers do exhibit a charged surface (e.g. (001) surface) but 
just like type 1 surfaces bear no dipole moment in the unit cel 
in a direction perpendicular to the surface. Hence, the surface 
energy is always finite.

What are the solutions to the polar catastrophe generated by a 
polar crystal truncation? 
As suggested by recent density functional calculations, a simple 
solution might prevail when the films are stoichiometric and 
only a few layers 
thick\cite{Goniakowski:2007p205701,Song:2008p035332}. 
The prescription is to reduce the dipole moment in each unit 
volume by essentially converting the two adjacent planes 
(e.g. Mg (2+) and O (2-) (111) planes in MgO) into one almost 
charge neutral plane with vanishingly small dipole. Similar 
behavior was found in the calculations of thin LaAlO$_3$ 
overlayers on a SrTiO$_3$ substrate\cite{Pentcheva:2009p107602}. 
These findings are consistent with the observation of a 
critical thickness in LaAlO$_3$/SrTiO$_3$ heterostructures for 
metallic behaviour of the interface\cite{Thiel:2006p1942}. 
There are, however, other plausible pathways whose 
contributions to the total energy of the system are independent 
of the film thickness. 

Basic electrostatic arguments suggest that the divergence of 
the electrostatic potential can be cured if the charge on the 
top and bottom surfaces is reduced to half of that of the 
corresponding layers in the bulk\cite{Kummer:1967p421}. 
Adsorption of a monolayer of foreign ions is, perhaps, the 
easiest way to achieve this without destroying the charge 
neutrality or introducing a significant structural modification 
on the surface. A notable example is a growth of $\alpha$-MnS 
single crystals using iodine as a transport agent in chemical 
transport method\cite{Heikens:1980p399}. The crystals are 
found to exhibit Mn terminated (111) faces with a monolayer 
of I$^-$ on top. Note, the crystal structure of $\alpha$-MnS 
is rock salt and hence this termination should be 
otherwise-unstable.  As another example, the hydroxyl groups 
have been observed on the polar surfaces of divalent oxide 
films\cite{Rohr1994L977,Cappus:1995p268,Tyuliev:1999p2900,
Lazarov:2005p115434,Ciston:2009p085421}.

The surface charge can also be altered by vacancies. 
For example, the recently discovered high temperature 
superconductor  BaFe$_2$As$_2$ has a nominal layered structure 
stabilized by the extremely large electronic polarizability of 
pnictogens\cite{Haas:1981p158,Wilson:1994p159,Sawatzky:2009p17006}.
It consists of negatively charged (001) FeAs sandwiches  
alternated with the Ba 1+ layers that generates a polar 
surface when the material is cleaved parallel to the layers.
Scanning tunnelling microscopy studies suggest that the 
surface generated by cleaving is the Ba-terminated surface 
with every other Ba missing\cite{Yin:2009p097002,Yin:2009p535}. 
A more involved defect structure is often discussed in relation 
to (111) surfaces of the ionic materials with a rock salt 
crystal structure. For example, the so-call octopolar 
reconstruction is one in which 3/4 of top layer atoms are 
removed together with 1/4 of those of the layer underneath, 
as shown in Fig.\ref{MgO111-octopolar}, creating again a 50\% 
charged surface\cite{Lacman:1965p195,Wolf:1992p16}. 
Interestingly, recent DFT studies of octopolar reconstruction 
on (111) surfaces of MgO and NiO suggest very strong deviation 
from a proposed structure toward rather flat surfaces 
consisting of a single buckled layer of metal and oxygen 
ions with a formal charge of $\pm1$\cite{Wander:2003p233405}.

Electronic reconstruction is yet another compensation mechanism. 
This mechanism reminds us of the Zener breakdown in which 
electronic tunnelling between two interfaces occurs when the 
external electric field exceeds the critical value imposed by 
the band gap of the dielectric\cite{Zener:1934p523}. There is, 
however, an important difference between Zener theory and polar 
catastrophe. In the latter case, the electric field is an 
intrinsic feature of an uncompensated film and should vanish 
if surface charges are exactly balanced with respect to charges 
of layers underneath.

\begin{figure*}
\includegraphics[clip=true,width=0.9\textwidth]{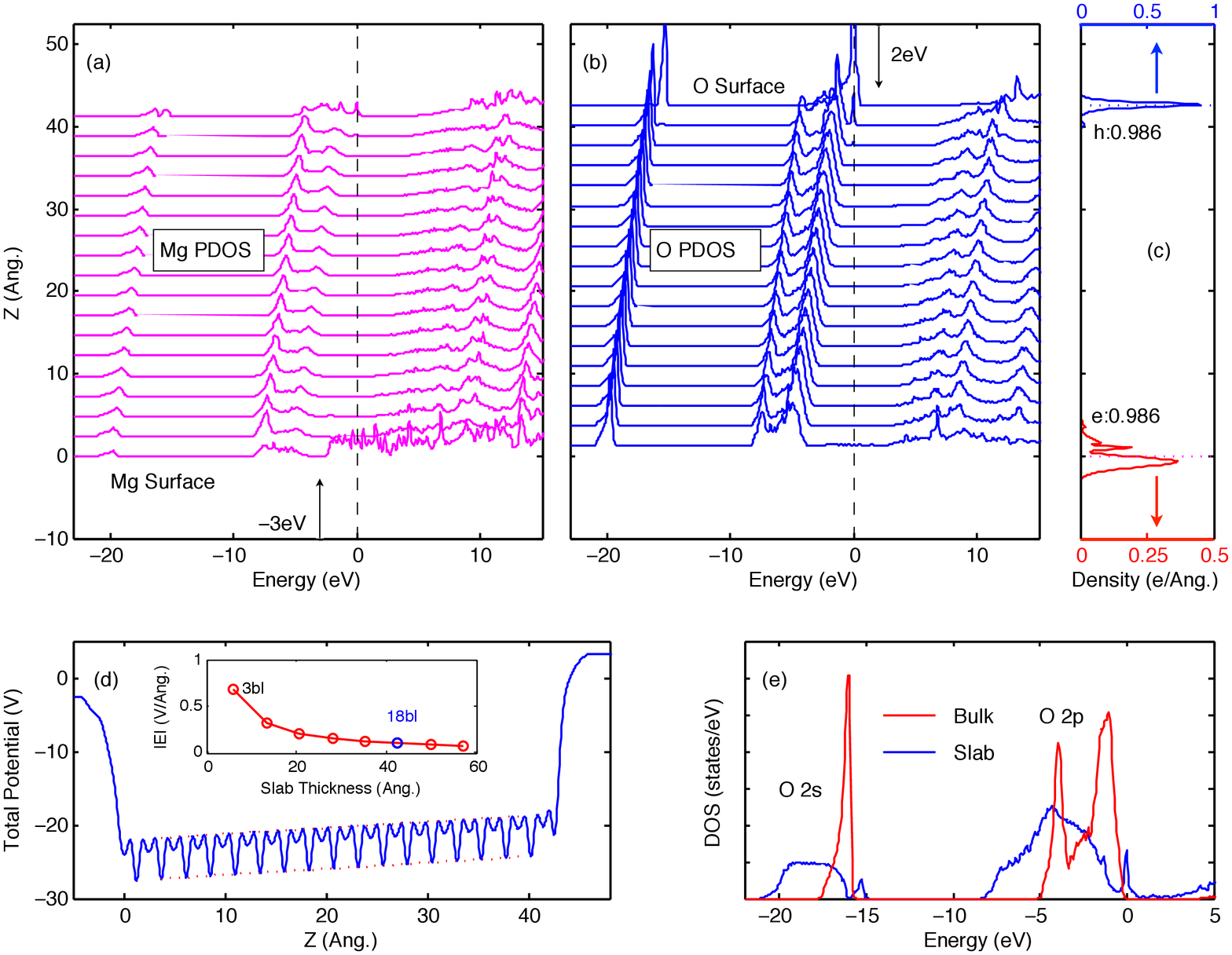}
\caption {GGA layer projected Mg (a) and O (b) density of 
states for 18 bilayer thick MgO (111) slab. The zero of energy 
is at the Fermi energy, $E_F$. The arrows depict the lower 
(electrons) and upper (holes) limits in the integration about 
$E_F$ used to calculate charge densities. Reconstructed charge 
distribution (c) and planar averaged total potential (d) in the 
slab as a function of distance z in direction parallel to 
surface normal. The residual electric field as a function of 
slab thickness (insert in (d)). A comparison between bulk and 
slab total density of states (e).}
\label{MgO111}
\end{figure*}

Let us consider a simple capacitor like model of a stack of $2N$ 
alternating charged planes with the charge density $\Sigma$ and 
separation $d$ as shown in Fig.\ref{geometry}(a).   
The potential difference between two interfaces is given by
\begin{equation}
V_{slab} = \frac {\Sigma - 2\sigma} {2\epsilon_0}(2N-1)d + 
\frac {\Sigma} {2\epsilon_0}d
\end{equation}
where $\sigma$ is the electron transfer between two interfaces 
($0 \leq \sigma \leq \Sigma$). Here, we refer to the surface as 
the interface.
The second term is the bulk electrostatic potential, $V_{bulk}$.
When $\sigma=0$, $V_{slab}$ is simply equal to twice the number 
of bilayers times the potential drop in the bulk material 
(e.g. potential drop between adjacent Mg and O (111) planes 
in the unit cell of MgO). To put it in prospective, the 4.3 
nm thick (111) film of MgO (18 bilayers) would experience 
an unsustainable 1030.7V potential across it. It is only 
about a factor of 2.4 smaller in LaAlO$_3$ (001) slab 
consisting of the same number of atomic layers (429.5V). 
Nevertheless, when the charge density on the both interfaces 
is reduced to a half ($\sigma=\frac{\Sigma}{2}$), the potential 
inside the slab equals the bulk potential and the electric 
field averages out to zero suggesting that the properties of 
fully charge compensated film are, interface layers aside, 
representative of that in the bulk. While capturing the basic 
electrostatics, such a simple model is lacking some key 
ingredients. Neglecting the polarizability effects, it is clear 
that redistribution of the electronic charge between two unlike 
interfaces requires an excitation across the band gap. 

This brings us to the main subject of the paper:  the role 
played by the band gap in electronic reconstruction of polar 
surfaces and interfaces. Let us define a quantity 
$\delta V = V_{slab} - V_{bulk}$ as a measure of the 
divergence in the electrostatic potential of the polar slab. 
As long as $\delta V$ is larger than the band gap,  the 
electrons would have to carry the compensating charge from 
one interface to another until $\delta V  = \Delta$ condition 
is fulfilled. Neglecting the band structure effects, the 
electron transfer is given by 
$q= \frac{1}{2} Q - \frac{\Delta \epsilon_0 A}{(2N-1)d}$, 
where $Q$ is a charge per unit area $A$ of the atomic plane 
in the bulk.
Note that $\Delta$ decreases the amount of charge transfer 
and only in the limit of very thick films 
($N \rightarrow \infty$) the films are fully electronically 
reconstructed ($q=\frac{1}{2} Q$). 
Using 7.8eV band gap and assuming fully ionic charges, we 
estimate the amount of charge left on the reconstructed 
surface of 18 bilayers thick MgO (111) slab to be $\pm$1.008e. 
This corresponds to a transfer of 0.992 electron between 
two surfaces of the slab. We get $\pm$1.004e surface charge 
when the gap is reduced to DFT value of 4.5eV. The incomplete 
reconstruction leaves behind the residual electric field 
of $\Delta/(2N-1)d$.  The field is greater for the larger 
band-gap materials and can alter the crystal structure of 
very thin films as was demonstrated recently by the DFT 
calculations of lattice relaxation in various polar 
films\cite{Pentcheva:2009p107602,Goniakowski:2007p205701,
Song:2008p035332,Wander:2003p233405}. Note that there is no 
internal field in the non-stoichiometric case.

Turning the discussion to the results of our DFT 
studies, we note that MgO has the rock salt crystal 
structure with a lattice constant of 4.213\AA~. It is 
a perfect dielectric with the band gap of 
7.8eV\cite{Roessler:1967p733,Whited:1973p1903}. 
Using experimental lattice constant, we get the band gap 
of 4.5 eV in agreement with previous density functional 
calculations\cite{Klein:1987p5802}. The MgO electronic 
structure is calculated using the pseudopotential DFT 
code SIESTA\cite{Soler:2002p2745} with the norm-conserving 
Troullier-Martins pseudopotentials\cite{Troullier:1991p1993} 
and the non-polar basis set of double-$\zeta$ quality 
for $s$-$p$ and single-$\zeta$ for $d$-orbitals. 
The exchange and correlation effects are treated within 
the generalized gradient approximation (GGA), after 
Perdew {\it et al}\cite{Perdew:1996p3865}. 
In the slab calculations, we adopt an ideal bulk crystal 
structure and  periodic slab geometry with 
100\AA~separation between the slabs. The effects of 
periodic boundary conditions will be discussed latter. 
The stoichiometric slabs consisting of 6-18 MgO bilayers 
are used throughout this study to address the polar 
catastrophe. No lattice relaxation has been performed in 
oder to isolate the effect due to electronic reconstruction.  

\begin{figure}
\includegraphics[clip=true,width=0.45\textwidth]{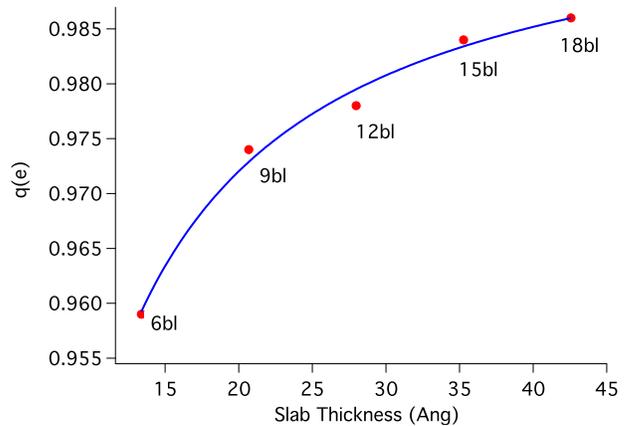}
\caption {Electron transfer (q) between two surfaces of the 
free standing MgO (111) film as a function of slab thickness 
obtained from DFT electron densities as described in the text. 
The solid line is a fit to the data using eq.\ref{eq_Q_mod}.}
\label{MgO111-q}
\end{figure}

Figures~\ref{MgO111} (a) and (b) show the layer projected 
density of states for 18 bilayers thick MgO film terminated 
with Mg layer on one side and Oxygen on the other.
The presence of the electric field is clearly evident from 
the shifts in the layer projected density of states and the 
slope of planar averaged total potential shown in 
Fig.~\ref{MgO111}(d). 
To find the exact amount of charge transfered between two 
surfaces, we calculate the two dimensional electron density 
of the slab as a function of the distance in a direction 
perpendicular to the slab. The electron and hole densities 
are calculated separately as the integrals over the ab-plane 
and over the energy from -3eV to $E_F$ (electron) and from 
$E_F$ to +2eV (hole) as depicted by the arrows in Fig.~\ref{MgO111} 
(a) and (b), respectively. To select the surface related 
contributions, the densities are truncated with a cutoff 
value of $10^{-4}$e/\AA. The apparent difference in spatial 
distribution of hole and electron densities as seen in 
Fig.~\ref{MgO111}(c) is due to the spatial extend and 
orientation of the atomic orbitals these particles are in. 
The electrons occupy rather extended and isotropic Mg 3$s$ 
orbitals whereas the holes are in in-plane O 2$p$ states. 
This is a result of the Hartree density-density repulsion 
between the O 2$p$ electrons which is at its minimum when 
the holes are confined to the surface.
Further integration over z reveals 0.9860$\pm 2\times10^{-4}$ 
electron transfer ($q$). It is remarkable how close it is to 
$\pm 1$ reconstruction expected for an ideal slab in which 
Mg and O are fully ionic with charges $\pm$2. 
Reduction in slab thickness results in a persistent decrease 
in $q$ as shown in fig.\ref{MgO111-q}. 
A deviation from the estimates given above can be attributed 
to actual electron density distribution and the band structure 
effects being not accounted for in the ionic model. 

As one can see in Fig.~\ref{MgO111}(c), the electron density 
on the Mg surface is rather extended in a direction away 
from the slab changing the electrostatic potential 
in the slab. Therefore, the expression for electron transfer 
needs to be modified. 
It is easy to show that, for this particular case, $q$ is given by
\begin{equation}
q = \frac{Q/2-\Delta \epsilon_0 A / D}{1+d_0/D}
\label{eq_Q_mod}
\end{equation}
where $D=(2N-1)d$ and $d_0$ is the distance from Mg surface to 
the mean of the electron density in vacuum. Note that the 
equation reduces to that given above in a limit of $d_0=0$. 
Using the 3.6V residual potential found in DFT calculation of 
18 bilayer thick slab, we get $Q=1.997\pm0.003$e and 
$d_0=0.39\pm0.03$\AA~by fitting eq.\ref{eq_Q_mod} to the DFT 
data for electron transfer obtained for various slab thicknesses.

Due to the residual electrostatic potential, the total density 
of states is broadened to, approximately, the total width of 
that in the bulk plus the gap energy. The effect is most 
evident in a case of O 2s states whose band widths are 
increased from 2eV in the bulk to about 5eV as shown in 
Fig.~\ref{MgO111}(e) suggesting that the spectroscopy of polar 
films could be very different from the bulk or films with non 
polar terminations. 

\begin{figure}
\includegraphics[clip=true,width=0.45\textwidth]{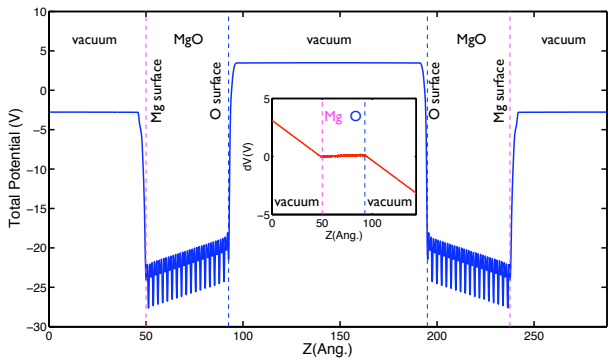}
\caption {MgO (111) planar averaged potential as a function 
of distance parallel to slab normal, $V(z)$, calculated in 
double slab geometry consisting of two slabs per supercell 
facing each other. The insert shows the difference with 
conventional slab model.}
\label{MgO-pot-comp}
\end{figure}

What about the periodic boundary condition? Since the slab 
is charge neutral there should be no electric field outside 
of the slab. The periodic boundary condition creates an artificial 
potential drop with the slope proportional to $1/D_V$ where 
$D_V$ is the distance between the slabs. This is analogous to 
non polar slabs with unlike terminations such as SrO and 
TiO$_2$ surfaces in SrTiO$_3$ (001) slab. This is often 
avoided by treating the surfaces separately e.g. SrTiO$_3$ 
(001) slab terminated with SrO layer on both sides. It is 
not an option in a case of polar slabs. One can wonder if 
an artifact due to PBC has any influence on the electronic 
structure of polar films. A simple way to resolve PBC 
problem is to create a double slab supercell with two slabs 
facing each other. In other words, the top surface of one 
slab is the same as the bottom surface of the slab above. 
The potential in vacuum is then constant which makes such 
a construction periodic. Fig.~\ref{MgO-pot-comp} shows a 
planar averaged potential as a function of distance in 
direction parallel to the surface normal calculated using 
double slab supercell. Apart from the vacuum region, there 
is no distinction between the potentials calculated using 
single and double slab supercells. In both cases, the 
residual potential inside the slab is about 3.6V. We 
conclude that for sufficiently large $D_V$ the electronic 
structure of the film calculated in a single slab geometry 
is not affected significantly by the periodic boundary condition. 

\begin{figure}
\includegraphics[clip=true,width=0.45\textwidth]{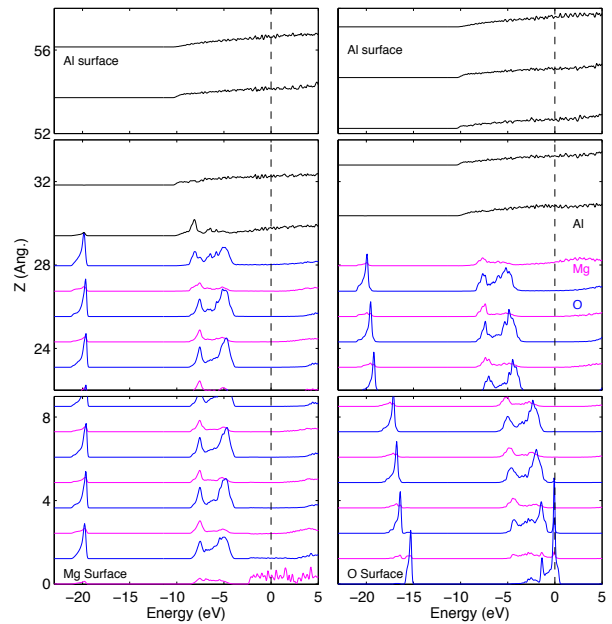}
\caption {GGA layer projected density of states for MgO (111) 
film on top of 18 monolayers of aluminum with O (left column) 
and Mg (right) contacts. Top, middle and bottom panels are Al 
surface, interface and MgO surface DOS, respectively. Color 
coding: Al is black, Mg and O are red and blue. The zero of 
energy is at the Fermi energy, $E_F$.}
\label{MgO-Al-dos}
\end{figure}

In free-standing films, the main cause for incomplete electronic 
reconstruction is the films band gap. It is, however, a conceptual 
system since films are never grown without the substrate. 
Furthermore, the interfaces are essential part of any 
heterostructure such as LaAlO$_3$/SrTiO$_3$. One can wonder 
if an interface with non-polar material can alter the electronic 
reconstruction of the polar films. Neglecting the band 
structure effects, this question leads to reexamination 
of the energy scales involved in the process. It is clear 
that, as long as the electronic structure on the non-polar 
side of the interface is such that it can provide states in 
the band gap of the polar one, it would have to be a part 
of the electronic reconstruction. Nevertheless, the charge 
redistribution within a single interface can not cure the 
polar catastrophe. Electrostatics requires that both interfaces 
participate in the same manner as in a free standing film. 

Let us now examine the electronic structure of a stoichiometric 
MgO (111) film on the surface of a metal, namely, aluminum. 
To keep the discussion coherent, we assume an ideal crystal 
structure and the lattice constant of MgO (Al is isotropically 
expanded). The lattice mismatch between Al and MgO is 3.9 percent. 
The structure is modelled by 18 bilayers of MgO and the same 
number of monolayers of Al in a single slab geometry with 
100\AA~of vacuum. The interface distances are chosen according 
to the minimum in the total energy as a function of MgO-Al 
distance. For the O-Al interface, we find that the separation 
is 1.44\AA~whereas it is substantially larger in a case of 
Mg-Al contact, 2.4\AA. 
Quick inspection of layer projected DOS shown in 
Fig.~\ref{MgO-Al-dos} reveals that
(i) the MgO surface projected DOS is very close to that 
in a free standing film (fig.\ref{MgO111});
(ii) the residual potential is 4.0V in MgO/Al slab 
containing Mg-Al interface and it is reduced to zero in 
the case of an O-Al contact layer; 
(iii) Al surface projected density of states is not 
affected by the interface with MgO.

\begin{figure}
\includegraphics[clip=true,width=0.45\textwidth]{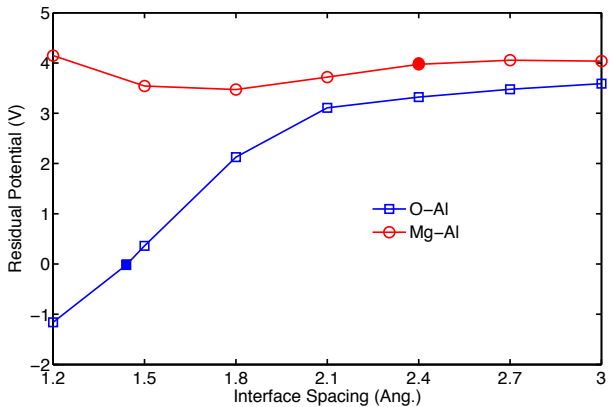}
\caption {The residual potential in MgO (111) film on the 
Al substrate as a function of interfacial spacing ($d_{int}$).
The solid symbols depict $d_{int}$ corresponding to a minimum 
in the total energy as a function of $d_{int}$.}
\label{MgO-Al-pot-distance}
\end{figure}

It is impossible to accurately predict the band alignment in 
the case of polar interface because the films are always 
reconstructed electronically in the calculations. 
However, the difference in $\delta V$ can be understood if 
the Fermi energy of Al was closer to the conduction band 
rather then the top of valence band of MgO. 
If that is the case, the redistribution of the charge 
between Mg surface and O-Al interface would necessitate 
the energy scale that is less than a half of the band gap 
which translates into smaller $\delta V$ and increases 
the electron transfer. This is exactly what we find. 
It is clear that having Al electronic states in the band 
gap of MgO helps to reduce the energy needed to reconstruct 
electronically because they provide the reservoir of the 
compensating charge at a fraction of the cost. The end 
result depends, however, on the film orientation (O vs Mg 
contact) and layer spacing at the interface. For example, 
we find that the total energy of the film with Mg-Al contact 
is about 3.1eV higher as compared to O-Al contact which is 
simply another evidence of much larger energy scale involved 
in the electronic reconstruction of the film. Again, we stress 
that due to electrostatic nature of the problem both surface 
and interface are strongly coupled via the electrostatic 
potential. This drives the electron transfer from Oxygen 
surface into Mg-Al interface that, in this particular 
configuration, costs almost as much as the band gap energy 
and leads to the residual potential which is nearly 
independent on the interface spacing, 
Fig.~\ref{MgO-Al-pot-distance}. The switch to another 
geometry allows to utilize much smaller energy scale and 
reduce the residual potential to zero at optimal O-Al distance. 
Of course, when this distance is large, the system is no longer 
coupled and the electronic reconstruction pathway is 
independent on the Al substrate. So, the reconstruction 
energy scale is, again, the band gap of MgO.

\begin{figure}
\includegraphics[clip=true,width=0.45\textwidth]{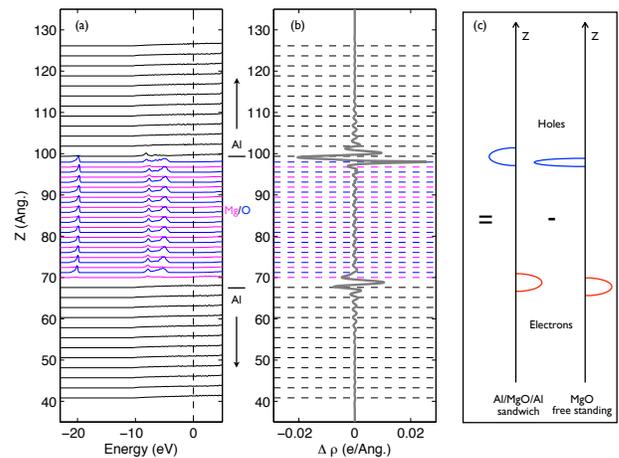}
\caption {Left panel - GGA layer projected DOS for 
Al$_{12}$(/MgO)$_{12}$/Al$_{12}$ slab. The zero of energy 
is at the Fermi energy. Middle panel - valence electron 
difference $\Delta \rho$ as a function of distance z in 
direction parallel to interface normal 
($\Delta \rho (z) = \rho_{Al/MgO/Al}^{slab} (z) - 
\rho_{Al}^{slab} (z) - \rho_{MgO}^{slab} (z)$). 
The oscillations inside MgO fragment of the structure 
are due to incomplete electronic reconstruction in free 
standing MgO (111) film. The oscillations at the interfaces 
are due to the change in reconstructed electron (hole) 
density distribution as compared to single MgO slab. 
Right panel - an illustration of the change in spatial 
distribution of compensating electron and hole densities 
due to interface with Al. The large oscillations at the 
interfaces in (b) are primarily due to readjustment of 
the compensating densities as compared free standing film.}
\label{Al-MgO-Al}
\end{figure}

To further demonstrate the interplay between the structure 
and the energy in electronic reconstruction pathways, we 
calculate the electronic structure of MgO (111) film sandwiched 
between two Al slabs (Al$_{12}$/(MgO)$_{12}$/Al$_{12}$). 
The interface distances are chosen to be equal to that found 
in MgO/Al slab calculations. Figure~\ref{Al-MgO-Al} shows the 
layer projected density of states and the planar integrated 
valence electron density difference defined as 
$\Delta \rho (z) = \rho_{Al/MgO/Al}^{slab} (z) - 
\rho_{Al}^{slab} (z) - \rho_{MgO}^{slab} (z)$. Comparing to 
Fig.~\ref{MgO-Al-dos}, it is clear that the MgO projected DOS 
is very similar to that in MgO/O-Al/Al slab with the exception 
of the Mg surface projected DOS. In particular, there is no 
shift in MgO DOS that suggests that the structure is fully 
compensated electronically. The electron density difference 
plotted in Fig.\ref{Al-MgO-Al}(b) as a function of distance z 
in a direction parallel to interface normal illustrates the 
role of Al band structure in the reconstruction pathway. As 
in other structures considered, the electron transfer is 
across the MgO slab but it is, to a large extent, Al density 
that participate in the process. This reduces the double 
interface energy ($E_{dint}$) to 1.92eV as compared to the 
slab surface energy ($E_{SSE}$) of 6.02eV in a free standing MgO. 
The $E_{dint}$ is given by
$$E_{dint} =  E_{tot}^{Al/MgO/Al} - N E_{tot}^{Al~bulk} - 
M E_{tot}^{MgO~bulk} - 2E_{surf}^{Al}$$
where $N$ ($M$) is the number of formula units in Al (MgO) slab; 
$E_{tot}$ is the respective total energy; $E_{surf}^{Al}$ is 
surface energy calculated for 12 monolayer thick Al (111) 
slab using MgO lattice constant (0.42eV). Correspondingly, 
the slab surface energy is 
$E_{SSE}=E_{tot}^{MgO~slab}-M E_{tot}^{MgO~bulk}$.

In conclusion, we demonstrate that the shifts in the DOS of 
polar systems are extremely sensitive to tiny deviations in 
electron transfer involved in the electronic reconstruction 
mechanism. They are not the signatures of electronic 
reconstruction and only tell us about its incompleteness. 
It is a result of an effective energy scale which is required 
in order for charge to be transferred between the oppositely 
charged interfaces. In a free standing film, this energy scale 
is set by the band gap modified to some degree to incorporate 
the band structure effects. At interfaces, it is determined by 
the band alignment and it is always the smallest energy possible 
helping the electronic reconstruction mechanism to be realized 
more readily in heterostructures as compared to free standing 
films. Due to high sensitivity of the inner potential variation 
to the interface charges, one can expect that the interface 
imperfections such as vacancies or adsorption of foreign species 
can reduce significantly the residual electrostatic potential even 
though the electron transfer energy scale of pure material is 
still rather large\cite{Gu:2010unpublished}. This can, in principle, 
reconcile theory and the results of x-ray photoelectron spectroscopy 
measurements in which no inner potential variation was 
found\cite{Heikens:1980p399,Sing:2009p176805,Kazzi:2006p954,
Hotta:2006p251916,Wadati:2008p045122,Takizawa:2009p236401}.
Furthermore, bulk impurity states energetically distributed 
across the band gap can somewhat modify the charge distribution 
at the interface in a fashion similar to conventional 
semiconductor P-N junctions.

We gratefully acknowledge Nicholas Ingle, 
Hiroki Wadati, Jochen Geck and Andrea Damascelli for stimulating 
discussions. This research was funded in part by the Canadian 
funding agencies NSERC, CRC, and CIFAR.

\bibliographystyle{apsrev}

\end{document}